\documentclass[aps,prl,twocolumn]{revtex4}

\usepackage{graphics}
\usepackage{epsfig}

\usepackage{color}

\def\avg#1{\left\langle#1\right\rangle}
\def\bra#1{\left\langle#1\right|}
\def\ket#1{\left|#1\right\rangle}

\begin{document}

\graphicspath{{Figures/}}

\title{General Relationship Between the Entanglement Spectrum and the Edge State
Spectrum of Topological Quantum States}

\author{Xiao-Liang Qi,$^{1,2}$ Hosho Katsura,$^{3,4}$ and Andreas W. W. Ludwig$^5$}
\affiliation{
$^1$Department of Physics, Stanford University, Stanford, CA 94305, USA\\
$^2$Microsoft Research, Station Q, University of
California, Santa Barbara, CA 93106, USA\\
$^3$Kavli Institute for Theoretical Physics, University of
California, Santa Barbara, CA 93106, USA\\
$^4$Department of Physics, Gakushuin University,
Mejiro, Toshima-ku, Tokyo 171-8588, Japan\\
$^5$Department of Physics, University of California, Santa Barbara, California 93106, USA
}

\date{\today}

\begin{abstract}
We consider (2+1)-dimensional  topological quantum  states which possess edge states
described by a chiral (1+1)-dimensional Conformal Field Theory (CFT), such
as e.g. a general quantum Hall state.  We demonstrate that for such states the
reduced density matrix of a finite spatial region of the  gapped topological state
is a thermal density matrix of the chiral edge state CFT
which would appear at the spatial boundary of that region.
We obtain this result by applying a physical instantaneous cut to the gapped system,
and by viewing the cutting process as a sudden ``quantum quench'' into a CFT,
using the tools of boundary conformal field theory.
We thus provide a demonstration of the observation made by Li and Haldane about the
relationship between the entanglement spectrum and the spectrum
of a physical edge state.
\end{abstract}


\maketitle


Topological phases of matter are gapped quantum states
which cannot be adiabatically
{
deformed into a
 completely
`trivial' gapped system
}
such as
a trivial band insulator,
without crossing a quantum phase transition.
They are not characterized by symmetry breaking,
but instead by certain global topological properties such as
the presence of (topologically)
protected edge states and/or a ground state degeneracy
which depends on the topology of the surface
on which the state resides\cite{WenNiuPRB1990}.
Topological states of matter
{
of this kind
which
}
have been discovered in nature
include the integer and fractional quantum Hall states\cite{PrangeGirvinBookQuanumHallEffect},
and the recently discovered time-reversal invariant topological insulators\cite{hasan2010,qi2010RMP,moore2009}.

Quantum entanglement is a purely quantum mechanical phenomenon which has no classical analog.
For any
{
pure quantum state (typically the ground state) of a system
}
consisting of two disjoint
subsystems $A$ and $B$,
{
subsystem $A$ can be described by a density matrix
obtained by `tracing out' the degrees of freedom in $B$.
}
This density matrix provides
complete information about the entanglement
{
properties of the initial pure state
}
{
between the two subsystems.
}
Quantum entanglement provides an alternative characterization of
the properties of the many-body system,
{
in particular for
}
topological states of matter which cannot be
{
described by conventional probes such as order parameters, correlation functions,
{\it etc.}. \cite{Amico}
}
For example, as discovered by Levin and Wen, and by Kitaev and Preskill\cite{LevinWen,KitaevPreskill}, the entanglement entropy of a topologically ordered state in a region of
{
linear size $l$
in two-dimensional position space
contains a universal $l$-independent term,
}
the  `topological entanglement entropy' (TEE), which
is a characteristic of the topological order of the state.
{
However, the TEE is not a complete description of a topological state of matter,
since
}
{
distinct
}
topologically ordered states can have the same TEE.
More complete information about a topological state of matter can be obtained from
the eigenvalue spectrum of the reduced density matrix,
often referred to as the entanglement spectrum\cite{Li07}. In general, the
density matrix $\rho_A$ describing the entanglement between a subsystem $A$ and the
rest of the system can be written in the form of $\rho_A=e^{-H_{\rm E}}$,
with $H_{\rm E}$ a Hermitian operator.
This is
{
so far
}
 nothing but the definition of $H_{\rm E}$.
In general, the
{
so-defined
``{\it entanglement Hamiltonian}" $H_{\rm E}$ is different from the
{\it physical Hamiltonian} of the system.
}
(In this form, the operator $H_{\rm E}$
{
appears formally  in a way analogous to
}
that of  the physical Hamiltonian $\beta H$
in a system in thermal equilibrium,
which has a thermal density matrix $\rho\propto e^{-\beta H}$
at temperature $T=1/k_{\rm B}\beta$.)
One important physical feature of the so-defined entanglement Hamiltonian $H_E$
is that its low-energy
eigenstates
correspond to those states in subsystem $A$,
appearing in the Schmidt-decomposition of the initial pure state, which are most entangled
with the rest of the system.

{
The focus of the present article is a remarkable observation
made recently by Li and Haldane\cite{Li07} and in subsequent work
for topological phases whose physical Hamiltonian possesses low energy states at an open
boundary (`edge states'). This includes
fractional quantum Hall states\cite{Li07, Laeuchli,Regnault10, Thomale10},
non-interacting topological insulators\cite{turner2009,Fidkowski09}
and the Kitaev honeycomb model\cite{yao2010}.
It was found that for those systems
the low-energy `edge' states of the physical Hamiltonian at an actual open
boundary of system A are in one-to-one correspondence with the
low-lying eigenstates of entanglement Hamiltonian $H_{\rm E}$ ({\it i.e.}
with the most entangled states).
However, except for systems which can be reduced to non-interacting fermion
problems\cite{turner2009,fidkowski2009,yao2010}, such a correspondence
}
between entanglement spectrum and the edge state spectrum of the physical Hamiltonian
has only been supported by numerical evidence.
No general argument for the validity of such a correspondence has
{
been presented so far\footnote{After our work was completed and while it was being written up, we
became aware of the preprint arXiv:1102.2218 in which general
analytical arguments were presented for the mentioned correspondence
in a large class of fractional quantum Hall states. The methods used
in both works are entirely different.}.
}
It is the purpose of the present Letter to demonstrate
the general validity of this correspondence.

\begin{figure}[htbp]
\centerline{\includegraphics[width=3.2in]{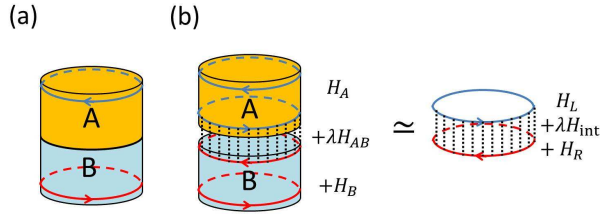}}
\caption{(a) A topological state on a cylinder with a bipartition into two regions $A$ and $B$. (b) The deformed system (see text) with the coupling between $A$ and $B$ regions weighted by a factor $\lambda\in[0,1]$. The system can be understood as two cylinders $A$ and $B$, with edge states propagating along the boundary between $A$ and $B$, coupled by an inter-edge coupling. (c) For small enough $\lambda$, the coupling between the gapped bulk states can be neglected, and the problem can be reduced to an inter-edge coupling problem described by a $(1+1)$-dimensional conformal field theory with a relevant coupling $\lambda H_{\rm int}$.}
\label{fig1}
\end{figure}

\noindent{\it General setup--}
In this Letter, we show that for a generic $(2+1)$-dimensional topological state which possesses
edges states described by a conformal field theory,
the entanglement Hamiltonian $H_{\rm E}$ is proportional to
{
the Hamiltonian $H_L$ of a physical chiral (say L-moving) edge
state appearing an actual spatial boundary of subsystem A,
in the long-wavelegnth limit and in  any fixed topological sector.
}
For example, our conclusion applies to all the Abelian and non-Abelian quantum Hall states
described by Chern-Simons effective field theories in the
bulk\cite{Zhang-Hansson-Kivelson, Blok-Wen, GeneralQuantumHallChernSimonsRef-Witten,
MooreRead}.
In order to relate the entanglement spectrum and
{
spectrum of the physical edge state Hamiltonian,
}
{
we consider a bipartition of the toplogical state on a  cylinder into two parts $A$ and $B$
}
as shown in Fig. \ref{fig1} (a).
{
The (physical) Hamiltonian
}
 $H$ can be written
{
in the form
}
\begin{eqnarray}
H=H_A+H_B+H_{AB}
\end{eqnarray}
where $H_A$ and $H_B$ denote the
Hamiltonians in
{
disconnected regions  $A$ and $B$, respectively,
each of which   has (two) open boundaries.
}
{
The term $H_{AB}$ couples regions $A$ and $B$
across their joint boundary.
}
For example, for a
2D gapped
tight-binding model
$H=\sum_{\avg{ij}}c_i^\dagger t_{ij}c_j$
realizing\cite{HaldaneModel} the
integer quantum Hall effect, the term $H_{AB}$ contains all the electron hopping terms across the boundary between
$A$ and $B$.

Now we consider a deformed Hamiltonian containing a parameter
$\lambda \in[0,1]$ {(similar to Ref. \cite{qi2005b})}:
\begin{eqnarray}
H(\lambda)=H_A+H_B+\lambda H_{AB}
\end{eqnarray}
By construction, $H(\lambda=0)$ is the Hamiltonian
of the two decoupled cylinders
$A$ and $B$, and $H(\lambda=1)$ is
the Hamiltonian of the whole cylinder $A \cup B$.
Since we are interested in such topological states which possess chiral edge states,
the Hamiltonian  $H(\lambda=0)$
will have chiral and anti-chiral edge states propagating  at the boundary
between regions $A$ and $B$, as shown in Fig. \ref{fig1} (b).
When $\lambda \neq 0$, the term $\lambda H_{AB}$ introduces a coupling between the
regions $A$ and $B$.
Denote the bulk gap of the
{
Hamiltonian $H= H(\lambda=1)$ by
}
$E_{\rm g}$. When the
coupling $\lambda$ is small enough such that the
energy scale of the coupling term $\lambda H_{AB}$ is much smaller than the bulk gap $E_{\rm g}$, the gapped bulk states described by $H_A$ and $H_B$ are almost entirely unaffected
by the coupling term $\lambda H_{AB}$, whose main effect is then to induce an inter-edge coupling between the chiral and anti-chiral edge states.
Since each individual edge state is described by a chiral conformal field theory (CFT), the theory of the two edges between regions $A$ and $B$
is described by a non-chiral conformal field theory. Thus at low-energy the coupling term $\lambda H_{AB}$ between regions $A$ and $B$
is reduced to a local interaction
in the CFT describing the dynamics of the two coupled edges.
If this interaction is a
relevant perturbation of the CFT describing the decoupled edges,
then the two counterpropagating edges
will be gapped for arbitrarily small coupling $\lambda$. Thus we expect that in this case the system described by the Hamiltonian $H(\lambda=1)$ to be
adiabatically connected to that described by $H(\lambda)$ for a small but non-vanishing
value of $\lambda$.
In this case, the entanglement properties of $H(\lambda =1)$
are expected to be qualitatively the same as those of $H(\lambda)$ with a small $\lambda$.
The latter describes the entanglement between the left- and the right-movers
{
of the edge state CFT.
Let us assume for simplicity that this coupling is relevant in the renormalization group (RG) sense.
}
\footnote{Even if the coupling term $\lambda H_{AB}$ is not relevant, our result still holds.
More details about that case is given in the supplementary material.}

Below we will solve this
entanglement
{
problem for the edge state CFT, by
}
mapping it to a  problem of a quantum quench.
We then solve the latter
{
(quantum quench)
}
 problem in the standard manner
by using
{
the work of Calabrese and Cardy\cite{calabrese2006, calabrese2007}
which employs the methods of boundary conformal field theory
(BCFT)\cite{Cardy_BCFT}.
}

\noindent{\it~Reduced density matrix of the edge CFT--}
{
Next we study
}
 the entanglement properties of the Hamiltonian
$H(\lambda)$
for small values of $\lambda$,
which, as explained above,  amounts  to the study of
the  $(1+1)$ dimensional problem of coupled edge states,
\begin{eqnarray}
H_{\rm edge}(\lambda)=H_L+H_R+ \lambda H_{\rm int}
\label{Hedge}
\end{eqnarray}
Here,
$H_L$ and $H_R$ denote the Hamiltonians of left-moving (L) and right-moving (R) edge states,
and $\lambda H_{\rm int}$
a {relevant} inter-edge coupling.
The left- and right- moving edge states are the low-energy excitations of
the subsystem in regions $A$ and $B$, respectively.
Again,
the entanglement
{
properties
 between
the subsystems $A$ and $B$
are
reduced to those
}
{
between left  and right  moving $(1+1)$-dimensional edge states.
}
If we denote the ground state of the
{
Hamiltonian $H_{\rm edge}(\lambda)$ from Eq.(\ref{Hedge})
by $G$,  then
}
our goal is to obtain the density matrix of the left-moving
{
edge state
}
subsystem defined by
\begin{eqnarray}
\rho_L={\rm Tr}_R\left(\ket{G}\bra{G}\right),
\end{eqnarray}
where ${\rm Tr}_R$
{
denotes
 the trace over the right-moving  edge state degrees of freedom.
}
In general, the ground state $\ket{G}$ will depend on all the details of the coupling
between the right- and the left- moving edges states.
However, due to the gapless nature of
$H_{\rm edge} (\lambda =0)$ describing the {\it decoupled} edges,
certain universal properties can be inferred
{
without reference to any detailed features of this coupling
in the long-wavelength limit.
}


In order to understand
the entanglement properties
of the state $\ket{G}$, we relate them
to another problem -- the ``quantum quench'' problem.
Consider a ``quantum quench'' of the system composed of the coupled
edges, Eq. \ref{Hedge}. For all times $t<0$ the system is in the
ground state $\ket{G}$  of the Hamiltonian  $H_{edge}(\lambda_0)$
with non-vanishing coupling $\lambda_0 \not =0$ between the edges. At time $t=0$
the coupling $\lambda_0$ between the edges is suddenly switched off, so that $\lambda=0$ for $t\geq 0$.
After the quantum quench, the left and right moving edge states evolve independently with the Hamiltonian
$H_{\rm edge}(\lambda=0)=$ $H_L+H_R$ of the decoupled edges. Space- and time-dependent
correlation functions after a sudden quench, as above, have been studied extensively
by Calabrese and Cardy\cite{calabrese2006,calabrese2007},
who applied
BCFT to obtain general
properties of such correlation function in the long-time and long-wavelength regime.
This is relevant for our purpose
{
because
the density matrix $\rho_L$ is uniquely determined by the set of
all the equal-time correlation functions of operators with support
solely on the left-moving edge,
}
\begin{eqnarray}
&&C(t,\left\{x_i\right\})=\nonumber\\
&&\bra{G}e^{it(H_L+H_R)}
\hat{O}_{L,1}(x_1) ...  \hat{O}_{L,n}(x_n)
e^{-it(H_L+H_R)}\ket{G}
\equiv
\nonumber \\
&&\equiv
{\rm Tr}_L\left[e^{-itH_L}\rho_Le^{itH_L}
\hat{O}_{L,1}(x_1) ...  \hat{O}_{L,n}(x_n)
\right]\label{correlation},
\end{eqnarray}
(All the coordinates $x_1,x_2,...x_n$
reside entirely on the left-moving edge.)
In the quantum quench problem, the ground state $\ket{G}$  of the coupled edge
Hamiltonian $H_{edge}(\lambda_0\not =0)$
{
represents
}
an initial condition at time $t=0$ for the evolution
{
with
}
the gapless (critical) decoupled
edge system
{
Hamiltonian $H_L + H_R$
}
at subsequent times $t>0$.
{
This initial state can be viewed\cite{Symanzik,calabrese2006,calabrese2007}
}
as a boundary condition on the gapless
{
theory of the right and left moving edges.
It can thus be described using the
methods of boundary critical phenomena\cite{diehl1986}.
This can be  achieved in any dimension of space
}
by analytical continuation of the real-time Keldysh
contour to imaginary time, and subsequent exchange of the roles of space and imaginary time (possible
due to the underlying effective relativistic invariance of the low-energy edge state theory).
Moreover, in the present case of a one-dimensional edge, the resulting boundary condition
can be analyzed by using the powerful tools of
BCFT\cite{Cardy_BCFT}.
The key result that we will use from the theory of boundary critical phenomena
is that an {\it arbitrary} boundary condition on a gapless bulk theory will
{\it always} renormalize at long distances into a
{
scale invariant boundary condition\cite{AffleckLudwigKondo,LudwigKondoReview,calabrese2006}.
}
Moreover, in the case of a (1+1) conformal bulk theory, such as the one
describing the one-dimensional edges, any scale invariant boundary conditions
must be one of a  known list of conformally invariant boundary conditions\cite{Cardy_BCFT}.
{
Consequently, as emphasized in
Ref. \cite{calabrese2006},
}
in the long wavelength limit the correlation functions
{
at a general boundary condition described by a
}
general state $\ket{G}$ are
equal to those
{
at a conformally invariant boundary condition described by a state $\ket{G_*}$ which
represents a (boundary)
fixed point to which the boundary state $\ket{G}$  flows under the renormalization group (RG).
}
The difference between $\ket{G}$ and $\ket{G_*}$
can be represented by an imaginary time evolution operator,
\begin{eqnarray}
\ket{G}\simeq Z^{-1/2}e^{-\tau_0(H_L+H_R)}\ket{G_*}\label{GandG0}
\end{eqnarray}
where $\tau_0>0$ is the so-called extrapolation length\cite{calabrese2006,diehl1986} and
stands for the RG ``distance"
of the general boundary state $\ket{G}$ to the conformal boundary state $\ket{G_*}$.
$Z=\bra{G_*}e^{-2\tau_0(H_L+H_R)}\ket{G_*}$ is a nomalization factor.
Physically, the energy scale
$1/\tau_0$ is determined by the energy gap $E(\lambda_0)$ induced by the coupling term $\lambda_0 H_{\rm int}$
between the edges.

{
In a so-called rational CFT such as the one under consideration, all
conformal invariant boundary states $\ket{G_*}$ are
known\cite{Cardy_BCFT}
to be
{\it finite} linear combinations of
so-called
Ishibashi states\cite{Ishibashi}
which have the form
}
\begin{eqnarray}
&&\ket{G_{*,a}}=
\nonumber\\
&&=\sum_{n=0}^{\infty}
\sum_{j=1}^{d_a(n)}
\ket{k(a,n),j;a}_L \otimes \ket{-k(a,n),j;\bar{a}}_R.
\label{Ishibashi}
\end{eqnarray}
{
Here $a$ denotes a topological sector in the underlying topological theory,
i.e. a topological flux threading the cylinder in Fig.
(\ref{fig1}(a)), which is
}
represented in the  CFT describing the edges
by a primary state of a corresponding conformal symmetry algebra (Virasoro or other)
of conformal weight $h_a$. ($\bar{a}$ denotes the conjugate sector and state of conformal weight
$h_{\bar{a}}=h_a$.)
{
The label $a$ runs over all possible particle types of the topological
state \cite{MooreSeiberg}.
}
Here $k(a,n)=2\pi(h_a+n)/l$ denotes the momentum, where
where $l$ is the circumference
of the edge of the cylinder;
$j=1,2,..d_a(n)$ labels the elements of an orthonormal basis
in the subspace of fixed  momentum $k(a,n)$.
Notice that the
{
left- (right-) moving edge system
}
only contains excitations
with positive (negative) momentum.
We note that
the state in Eq. \ref{Ishibashi} is and example of a so-called {\it maximally entangled}
state. The explicit form, Eq. \ref{Ishibashi}, of the Ishibashi states $\ket{G_{*,a}}$, resulting from conformal invariance,
is of great help in determining the form of the reduced density matrix
$\rho_L$ for the left-moving edge.
{
Upon directly combining Eqs. (\ref{GandG0}) with (\ref{Ishibashi})
one obtains
}
\begin{eqnarray}
&&\ket{G_a}\simeq
\nonumber \\
&& \simeq
\sum_{n=0}^\infty
{
e^{-2\tau_0 v k(a,n) }
\over Z^{1/2}}
\sum_{j=1}^{d_a(n)}\ket{k(a,n),j;a}_L
\otimes \ket{-k(a,n),j;\bar{a}}_R
\nonumber
\end{eqnarray}
{
which
yields the following form of the density matrix of the left-moving
edge upon tracing out the right moving edge,
}
\begin{eqnarray}
&&\rho_{La}={\rm Tr}_R\left(\ket{G_a}\bra{G_a}\right)\simeq
\nonumber\\
&&\simeq
\sum_{n=0}^\infty
{ e^{-4\tau_0v k(a,n) } \over Z }
\sum_{j=1}^{d_a(n)} \ket{k(a,n),j;a}_L\bra{k(a,n),j;a}_L =
\nonumber\\
&&
=
Z^{-1}
{\hat P}_a
\ e^{-4\tau_0H_L}
{\hat P}_a
\label{entanglement}
\end{eqnarray}
{
Here we have used the linear dispersion $H_L\ket{k,j;a}_L=v k(a,n)\ket{k,j;a}_L$,
$H_R\ket{-k,j;\bar{a}}_L=v k \ket{-k,j;\bar{a}}_R$
where $v$ is the edge state velocity and $k$ stands for $k(a,n)$.
The label $a$ indicates
that $\rho_{La}$ is an operator defined
in the topological sector corresponding to topological flux $a$
threading the cylinder,
and ${\hat P}_a$ is the projection operator onto that sector of the Hilbert space of the CFT.
}
In cylinder geometry there is no entanglement between different topological sectors
{
(denoted by different labels $a$).
}

Eq. (\ref{entanglement}) is the central result of this work, which demonstrates that the entanglement
between left-moving and right-moving edge states in a CFT
induced by a relevant coupling is always characterized by a ``thermal" density matrix
within a fixed topological sector
{
(or primary state, in the CFT context).
}
In other words, in each topological sector
the ``entanglement Hamiltonian" $H_{\rm E}=-\log\rho_L=4\tau_0H_L+\log Z$
{
is proportional to the
Hamiltonian $H_L$ of a physical edge
}
up to a possible shift of the ground state energy in that sector
which ensures the proper normalization of the density matrix as a probability
distribution.
{
Our result demonstrates not only that the excitation energies of the entanglement
spectrum are the same as those of the spectrum of the
Hamiltonian of the edge state of the topological system
appearing (by assumption) at a physical boundary of region A,
in the long-wavelength limit modulo a global rescaling, but also that the most
entangled states are in one-to-one correspondence with the
low-energy edge states which occur at this boundary.
}

\noindent{\it~Example: Free fermions--}
A simple example in which the general notions, developed in the preceeding part of
this article, can also be illustrated using elementary many-body techniques
is that of the 2D integer quantum  Hall (IQH) state (and more generally
non-interacting topological insulators).
This state can be described by a free fermion theory,
{
the entanglement properties of which
have been studied extensively
}
 in the literature\cite{Peschel03, turner2009, fidkowski2009}.
However, it is still helpful to present the results here as an illustration,
{
in the language of the much more general formulation obtained above.
}
The edge states of an IQH state with
{
integer
}
filling fraction $\nu=N$ consist of $N$
flavors of non-interacting chiral fermions. For simplicity, we consider an IQH state
with filling fraction $N=1$,
{
whose edge state dynamics is governed by the Hamiltonian
}
\begin{eqnarray}
H_L=\sum_k vkc_k^\dagger c_k,~H_R=-\sum_kvkd_k^\dagger d_k\label{HfreeCFT}
\end{eqnarray}
The simplest inter-edge coupling term is a single-particle inter-edge tunneling
\begin{eqnarray}
H_{\rm int}=E_{\rm g}\sum_k\left(c_k^\dagger d_k+d_k^\dagger c_k\right)\label{Hfreeint}
\end{eqnarray}
with $E_{\rm g}$ the bulk gap which acts as a
{
high-energy
}
cut-off
scale for the edge theory. The coupled Hamiltonian $H_L+H_R+\lambda H_{\rm int}$ is a free Fermion Hamiltonian
which can be diagonalized by a unitary transformation
to $H_L+H_R+H_{\rm int}=\sum_{k,s=\pm1}E_{k}\gamma_{ks}^\dagger\gamma_{ks}$
with the gapful
energy dispersion $E_{k}=\sqrt{v^2k^2+E_{\rm g}^2}$. Here $\gamma_{k,i}$ ($i=1,2$)
are quasiparticle annihilation operators.
The ground state $\ket{G}$ of this gapped system is determined by the
conditions $\gamma_{k, i}\ket{G}=0$ ($i=1,2$).
{
One obtains\footnote{Details are provided in the Supplementary Material} the
following explicit expression for $\ket{G}$ (unnormalized):
}
\begin{eqnarray}
\ket{G}&=&e^{-H_{\rm e}}\ket{G_*}\label{GandG0free}
\end{eqnarray}
with
\begin{eqnarray}
\ket{G_*}&=&\exp\left\{-\sum_{k>0}\left(c_k^\dagger d_k+d_{-k}^\dagger c_{-k}\right)\right\}\ket{G_{L}}\otimes\ket{G_{R}}.
\nonumber\\
\end{eqnarray}
and $H_e\simeq \frac1{2E_{\rm g}}(H_L+H_R)$ in the long wavelength limit.
{
The operators $c^\dagger_k d_k$ and $d^\dagger_{-k} c_{-k}$ with $k>0$ create quasiparticle
excitations of the system of the two edges, so that $\ket{G_*}$
}
is an equal-weight superposition of all quasi-particle excitation states in the massless
{
theory; this is nothing but the Ishibashi state for the Free fermion CFT (in the sector without topological
flux).
}
{
Thus, with this form of $H_e$,
we recover correctly (in the long wavelength limit) the general relation (\ref{GandG0});
the extrapolation length is $\tau_0=1/2E_{\rm g}$.
}
As expected, the energy scale $1/\tau_0$ is
determined by the energy gap $2E_{\rm g}$ of particle-hole excitations.

\noindent{\it~Conclusion and discussion--}
In conclusion, we have demonstrated that for a generic $(2+1)$ dimensional topological state
possessing gapless edge states
{
which are
}
described by a chiral CFT, the reduced density matrix of
of a region $A$ obtained by tracing out the rest of the system always has,
in the long wavelength limit,  the same
{
form as a thermal density matrix of region $A$ with an open
physical
} boundary.
Besides topological states, our analysis also applies to other systems described by coupled CFTs.
In particular, our result
{
provides an explanation
of the
}
recent numerical and analytical results on the entanglement spectrum of coupled spin chains\cite{Poilblanc}
and coupled Luttinger liquids\cite{Furukawa_Kim}.
Since the relationship between a general boundary state and
a scale invariant boundary condition which is the endpoint of the RG flow
also holds for higher dimensional
scale invariant bulk theories\cite{diehl1986},
we expect that our result
will generalize to higher dimensional topological states,
such as $(3+1)$ dimensional topological insulators,
{
and especially
}
 the fractional topological insulators
{
\cite{
Stern09,
maciejko2010b,swingle2010,cho2010}
}
which cannot be analysed using free fermion methods\cite{turner2009,fidkowski2009}.
Details of this generalization will be left for future work.

We would like to note that the reduced density matrix (\ref{entanglement})
in the topological sector $a$
{
yields
}
 an entanglement entropy of the form $S=-{\rm Tr}\left(\rho_L\log \rho_L\right)=\alpha L-S_{\rm topo}$,
with $S_{\rm topo}=\log \left(D/d_a\right)$ the topological entanglement entropy\cite{LevinWen,KitaevPreskill}.
Here $d_a$ is the quantum dimension of the quasi-particle
of type $a$,
and $D=\sqrt{\sum_a d_a^2}$ is the
 total quantum dimension. This relation
to the topological entropy
has been noticed in
Ref. \cite{KitaevPreskill},
{
though in that work the
 form (\ref{entanglement}) of the
density matrix was taken as an assumption.
}
The present paper proves this assumption.


HK was supported in part by the National Science Foundation under Grant
No. PHY05-51164 and JSPS.
This work was supported, in part, by the NSF under Grant No. DMR-
0706140 (A.W.W.L.), Alfred P. Sloan Foundation (X.L.Q.).

\bibliography{topo_entropy,TI}

\begin{widetext}
\appendix
\section{Supplentary Material I: More detailed discussion on irrelevant inter-edge coupling}

In the main text we have mainly studied the cases in which the interaction term $\lambda H_{\rm int}$ in Eq. (\ref{Hedge}) is a relevant coupling in the edge CFT, so that a gap is induced once a finite coupling $\lambda\neq 0$ is turned on. In this appendix, we will discuss the cases in which the interaction term $\lambda H_{\rm int}$ is irrelevant, and provide arguments that our result on entanglement spectrum still holds in this case.

For concreteness of the discussion, we study the simplest fractional quantum Hall state--$1/m$ Laughlin state as an example. In the cylinder geometry shown in Fig. \ref{fig1}, the two edges in the middle are described by a Luttinger liquid
(see e.g. \cite{XGWen_chiral_boson})
\begin{eqnarray}
\mathcal{L}_0=\frac{m}{2\pi}\left(\partial_t-v\partial_x\right)\phi_L\partial_x\phi_L+\frac{m}{2\pi}\left(-\partial_t-v\partial_x\right)\phi_R\partial_x\phi_R\label{FQHedge}
\end{eqnarray}
The inter-edge interaction can be written as\cite{moon1993}
\begin{eqnarray}
\mathcal{L}_{\rm int}&=&\lambda\cos\left[\frac1R(\phi_L-\phi_R)\right]\label{FQHint}
\end{eqnarray}
Physically, electron inter-edge tunneling corresponds to $R=1$. When a forward scattering term such as $g\partial_\mu\phi_L\partial^\mu\phi_R$ is turned on together with the back-scattering term, the compactification radius $R$ can deviate from $1$. For fractional quantum Hall edge states $m>1$ is an odd integer, and the coupling
$\lambda$ is irrelevant if $R=1$, as shown in Fig. \ref{fig2}. If $\lambda$ is continuously tuned starting from $\lambda=0$, the system remains gapless until $\lambda=\lambda_c$ where a phase transition occurs and the system becomes gapped. By construction, the physical system (Fig. \ref{fig1} (a)) without edge between A and B region corresponds to some value $\lambda>\lambda_c$ in the gapped phase, as shown in Fig. \ref{fig2} by point A. For $\lambda<\lambda_c$ the coupling is irrelevant, but the compactification radius $R$ of the theory is renormalized. In other words, for $\lambda<\lambda_c$ the long wavelength behavior of the coupled system is still described by a CFT but it is different from the original CFT in Eq.(\ref{FQHedge}). Thus naively it seems that our derivation in the main text does not directly apply to this theory.  
However, if we also turn on the forward scattering which increases $R$, the scaling dimension of $\lambda$ can be tuned to the region where $\lambda$ is relevant, as shown in Fig. \ref{fig2} by point B. At this point Eq. (\ref{entanglement}) applies, and the reduced density matrix can be written as $\rho_L=Z^{-1}e^{-4\tau_0 H_L}$ in each given topological sector. Now consider a path in the parameter space connecting A and B parameterized by $(\lambda(t),R(t))$ with $t\in[0,1]$. As long as the path stays in the gapped phase, for each point on the path the system has a gapped ground state $\ket{G(\lambda(t),R(t))}$, which determines a reduced density matrix $\rho_L(\lambda(t),R(t))={\rm Tr}_R\ket{G(\lambda(t),R(t))}\bra{G(\lambda(t),R(t))}$ correspondingly. Since no phase transition occurs along the path, the wavefunction of the state $\ket{G(\lambda(t),R(t))}$ is a smooth function of $\lambda$ and $R$, so that $\rho_L(\lambda(t),R(t))$ is also smooth. Consequently, there is a one-to-one correspondence between the states in the low-lying entanglement spectrum of $\rho_L$ at the two points A and B. In other words, the low-lying entanglement spectrum of point A contains the same states as the chiral CFT of the edge theory. The eigenvalues of $H_E=-\log \rho_L(\lambda(t),R(t))$ can change continuously during the deformation, but the long-wavelength limit of the dispersion relation has to remain linear since the deformation is smooth. Thus in the long wavelength limit, the reduced density matrix $\rho_L$ at A point still has the form of Eq. (\ref{entanglement}), with generically a different $\tau_0$ from B point.

\begin{figure}[htbp]
\centerline{\includegraphics[width=3.2in]{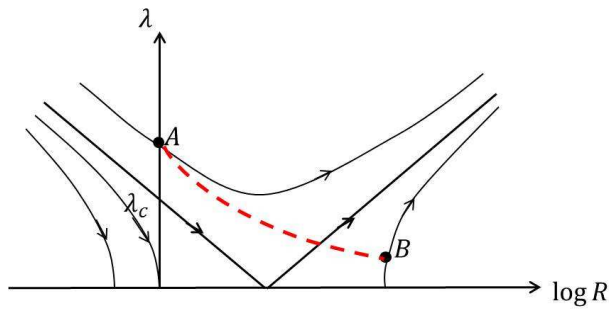}}
\caption{Illustration of the RG flow of model (\ref{FQHedge}) and (\ref{FQHint}) with two parameters $\lambda$ and $R$. Any two points A and B in the gapped phase can be connected by a continuous path (red dash line). By this continuous deformation one can show that the entanglement spectrum of system A is qualitatively the same as system B. The entanglement Hamiltonian of system B can be obtained using the approach in the main text since $\lambda$ is relevant.}
\label{fig2}
\end{figure}

The argument above applies to more generic CFTs as long as such path in the parameter space of the theory exists, which smoothly connects the parameter region where the coupling is irrelevant to the region where the coupling is relevant. In other words, when the coupling $\lambda H_{\rm int}$ is irrelevant, our conclusion on the relation between entanglement spectrum and edge state spectrum still holds {\it as long as there is a marginal coupling in the CFT which can tune the scaling dimension of $H_{\rm int}$ continuously to make it relevant.} 

\section{Supplementary Material II: Detailed derivation of the free Fermion density matrix}

The free fermion Hamiltonian given by Eq. (\ref{HfreeCFT}) and (\ref{Hfreeint}) can be diagonalized in the following form:
\begin{eqnarray}
H_L+H_R+\lambda H_{\rm int}&=&\sum_{k,s=\pm1}E_{k}\gamma_{ks}^\dagger\gamma_{ks}\\
\text{with~}E_{k}&=&\sqrt{v^2k^2+E_{\rm g}^2},~\left(\begin{array}{c}\gamma_{k+}\\\gamma_{-k,-}^\dagger\end{array}\right)=\left(\begin{array}{cc}u_k&-v_k\\v_k&u_k\end{array}\right)
\left(\begin{array}{c}c_k\\d_k\end{array}\right),~
\left(\begin{array}{c}u_k\\v_k\end{array}\right)=\left(\begin{array}{c}\sqrt{\frac{E_k+ vk}{2E_k}}\\-\sqrt{\frac{E_k- vk}{2E_k}}\end{array}\right)\nonumber
\end{eqnarray}

The ground state $\ket{G}$ of this gapped system can be determined by the conditions $\gamma_{ks}\ket{G}=0$ for $s=\pm 1$, which leads to the following form (not normalized)
\begin{eqnarray}
\ket{G}&=&\exp\left\{\sum_{k>0}\frac{v_k}{u_k}c_k^\dagger d_k+\sum_{k<0}\frac{u_k}{v_k}d_k^\dagger c_k\right\}\ket{G_{L}}\otimes\ket{G_{R}}\nonumber\\
\end{eqnarray}
with $\ket{G_{L(R)}}$ the ground states of the gapless systems $H_L$ and $H_R$, respectively. The operators $c_k^\dagger d_k$ and $d_k^\dagger c_k$ creates particle-hole excitations in the massless theory. The physical meaning of this state can be understood better by redefining the quasi-particle creation operators in the massless theory as $L_{1k}=c_k$ and $L_{2k}=c_{-k}^\dagger$ for $k>0$, and similarly $R_{1k}=-d_{-k}^\dagger$ and $R_{2k}=d_{k}$ for $k<0$. In term of these operators,
\begin{eqnarray}
\ket{G}&=&\exp\left\{\sum_{k>0}\frac{E_{\rm g}}{E_k+vk}\Delta_k^\dagger\right\}\ket{G_{L}}\otimes\ket{G_{R}}
\end{eqnarray}
with
\begin{eqnarray}
\Delta_k^\dagger=L_{k1}^\dagger R_{-k1}^\dagger+L_{k2}^\dagger R_{-k2}^\dagger=-\left(c_k^\dagger d_k+d_{-k}^\dagger c_{-k}\right)
\end{eqnarray}
 creates a pair of quasi-particles in the two edges. Defining
\begin{eqnarray}
H_e&=&-\frac12\sum_{k>0,s=1,2}\log\left({\frac{E_{\rm g}}{E_k+vk}}\right)\left(L_{ks}^\dagger L_{ks}+R_{-k,s}^\dagger R_{-k,s}\right)\nonumber\\
&=&-\frac12\sum_{k>0,s=1,2}\log\left({\frac{E_{\rm g}}{E_k+vk}}\right)\left(c_k^\dagger c_k-d_k^\dagger d_k+c_{-k}c_{-k}^\dagger-d_{-k}d_{-k}^\dagger\right)\label{freeHE}
\end{eqnarray}
we have the identity
\begin{eqnarray}
e^{-H_e}L_{ks}^\dagger e^{-H_e}=\sqrt{\frac{E_{\rm g}}{E_k+vk}} L_{ks}^\dagger,~e^{-H_e}R_{ks}^\dagger e^{-H_e}=\sqrt{\frac{E_{\rm g}}{E_k+vk}} R_{ks}^\dagger
\end{eqnarray}
Consequently $\ket{G}$ can be rewritten as
\begin{eqnarray}
\ket{G}&=&e^{-H_e}\exp\left\{\sum_{k>0}\Delta_k^\dagger\right\}e^{-H_e}\ket{G_{L}}\otimes\ket{G_{R}}\nonumber\\
&=&e^{-H_e}\exp\left\{\sum_{k>0}\Delta_k^\dagger\right\}\ket{G_{L}}\otimes\ket{G_{R}}\equiv e^{-H_e}\ket{G_*}\label{GandG0free}
\end{eqnarray}
In the last line, we have used the fact that $H_e\ket{G_{L}}\otimes\ket{G_{R}}=0$. 
In the long-wavelength limit $k\rightarrow 0$, $H_e$ in Eq. (\ref{freeHE}) is expanded as
\begin{eqnarray}
H_e&\simeq&\sum_{k>0,s=1,2} \frac{vk}{2E_{\rm g}}\left(L_{ks}^\dagger L_{ks}+R_{-k,s}^\dagger R_{-k,s}\right)\nonumber\\
&\equiv&\frac1{2E_{\rm g}}\left(H_L+H_R\right)
\end{eqnarray}
Consequently, Eq. (\ref{GandG0free}) in the long wavelength limit correctly recovers the general relation (\ref{GandG0}) with the extrapolation length $\tau_0=1/2E_{\rm g}$.

\end{widetext}
\end{document}